\documentclass{PoS}
\newcommand{\p}{\partial}
\newcommand{\pslash}{p\kern-1ex /}
\newcommand{\lslash}{l\kern-1ex /}
\newcommand{\kslash}{k\kern-1ex /}
\newcommand{\dslash}{\p\kern-1.2ex /}
\newcommand{\Dslash}{{\cal D}\kern-1.5ex /}
\newcommand{\Aslash}{A\kern-1.2ex /}

\newcommand{\tr}{{\rm tr}}

\newcommand{\bea}{\begin{eqnarray}}
\newcommand{\eea}{\end{eqnarray}}
\newcommand{\vol}{\Omega}

\newcommand{\BAN}{\begin{eqnarray*}}
\newcommand{\EAN}{\end{eqnarray*}}
\newcommand{\NTU}{
  Physics Department, Center for Theoretical Sciences, 
  and Center for Quantum Science and Engineering,  
  National Taiwan University, Taipei~10617, Taiwan  
}

\newcommand{\Tsukuba}{
  Graduate School of Pure and Applied Sciences, University of Tsukuba,
  Tsukuba 305-8571, Japan
}

\newcommand{\BNL}{
  Riken BNL Research Center, Brookhaven National Laboratory, Upton,
  NY11973, USA
}

\newcommand{\KEK}{
  High Energy Accelerator Research Organization (KEK),
  Tsukuba 305-0801, Japan
}

\newcommand{\GUAS}{
  School of High Energy Accelerator Science,
  The Graduate University for Advanced Studies (Sokendai),
  Tsukuba 305-0801, Japan
}

\newcommand{\YITP}{
  Yukawa Institute for Theoretical Physics, 
  Kyoto University, Kyoto 606-8502, Japan
}

\newcommand{\RCAS}{
  Research Center for Applied Sciences, Academia Sinica,
  Taipei~115, Taiwan
}
\title{Topological susceptibility in (2+1)-flavor lattice QCD 
       with overlap fermion}

\ShortTitle{Topological susceptibility in (2+1)-flavor lattice QCD}

\author{\speaker{T.W.~Chiu}$^{,a}$\thanks{Email: twchiu@phys.ntu.edu.tw},
        S.~Aoki$^{b,c}$, 
        S.~Hashimoto$^{e,f}$,
        T.H.~Hsieh$^g$,
        T.~Kaneko$^{e,f}$,
        H.~Matsufuru$^{e}$,
        J.~Noaki$^{e}$,
        T.~Onogi$^{h}$,
        N.~Yamada$^{e,f}$ (for JLQCD and TWQCD Collaborations)   
        \\
         \llap{$^a$}\NTU      \\
         \llap{$^b$}\Tsukuba  \\
         \llap{$^c$}\BNL      \\
         \llap{$^e$}\KEK      \\
         \llap{$^f$}\GUAS      \\
         \llap{$^g$}\RCAS     \\
         \llap{$^h$}\YITP     \\
}

\abstract{
  We determine the topological susceptibility 
  $ \chi_t $ in the topologically-trivial sector 
  generated by lattice simulations of $ N_f = 2+1 $
  QCD with overlap Dirac fermion, on a $16^3 \times 48 $ lattice with
  lattice spacing $\sim$ 0.11 fm, for five sea quark masses $m_q$ ranging
  from $m_s/6$ to $m_s$ (where $m_s$ is the physical strange quark mass).
  The $ \chi_t $ is extracted from the plateau (at large time separation)
  of the 2-point and 4-point time-correlation functions of the
  flavor-singlet pseudoscalar meson $\eta'$, which arises from the
  finite size effect due to fixed topology.
  In the small $m_q$ regime, our result of $\chi_t $ agrees with 
  the chiral effective theory. 
  Using the formula $ \chi_t = \Sigma(m_u^{-1} + m_d^{-1} + m_s^{-1})^{-1} $
  by Leutwyler-Smilga, we obtain the chiral condensate 
  $\Sigma^{\overline{\mathrm{MS}}}(\mathrm{2~GeV})
  =[249(4)(2) \mathrm{MeV}]^3 $. 
}

\FullConference{The XXVI International Symposium on Lattice Field Theory\\
                July 14-19 2008\\
                Williamsburg, Virginia, USA}

\begin{document}

\section{Introduction}

In Quantum Chromodynamics (QCD), the topological susceptibility
($ \chi_t $) is the most crucial quantity to measure the
topological charge fluctuations of the QCD vacuum,
which plays an important role in breaking the $ U_A(1) $ symmetry.
Theoretically, $ \chi_t $ is defined as
\bea
\label{eq:chi_t}
\chi_{t} = \int d^4 x  \left< \rho(x) \rho(0) \right>, \hspace{4mm} 
\rho(x) = \frac{1}{32 \pi^2} \epsilon_{\mu\nu\lambda\sigma}
                             \tr[ F_{\mu\nu}(x) F_{\lambda\sigma}(x) ]
\eea
where $ \rho(x) $ 
is the topological charge density 
expressed in term of the matrix-valued field tensor $ F_{\mu\nu} $.
With mild assumptions, Witten \cite{Witten:1979vv} and
Veneziano \cite{Veneziano:1979ec}
obtained a relationship between the topological susceptibility
in the quenched approximation and the mass of $ \eta' $ 
meson (flavor singlet) in full QCD with $ N_f $ degenerate flavors, 
namely, $ \chi_t(\mbox{quenched}) = f_\pi^2 m_{\eta'}^2/(4 N_f) $
where $ f_\pi = 131 $ MeV, the decay constant of pion.
This implies that the mass of $ \eta' $ is essentially due to
the axial anomaly relating to non-trivial topological charge
fluctuations, which can turn out to be nonzero even in the chiral limit,
unlike those of the (non-singlet) approximate Goldstone bosons.

Using the Chiral Perturbation Theory (ChPT), 
Leutwyler and Smilga \cite{Leutwyler:1992yt}
obtained the following relation in the chiral limit  
\bea
\label{eq:LS}
\chi_t = \frac{\Sigma}
         {\left( \frac{1}{m_u} +\frac{1}{m_d} +\frac{1}{m_s} \right)}
+ {\cal O}(m_u^2), \hspace{4mm} (N_f = 2+1), 
\eea
where $ m_u$, $m_d$, and $m_s$ are the quark masses, 
and $ \Sigma $ is the chiral condensate.
This implies that in the chiral limit ($ m_u \to 0 $) 
the topological susceptibility is suppressed due to internal quark loops.
Most importantly, (\ref{eq:LS}) provides a viable way
to extract $ \Sigma $ from $ \chi_t $ in the chiral limit.

From (\ref{eq:chi_t}), one obtains
\BAN
\chi_t = \frac{\left< Q_t^2 \right>}{\Omega}, \hspace{4mm}
Q_t \equiv  \int d^4 x \rho(x), 
\EAN
where $ \Omega $ is the volume of the system, and
$ Q_t $ is the topological charge.
% (which is an integer for QCD).
Thus, one can obtain $ \chi_t $ by counting the number of
gauge configurations for each topological sector.
Obviously, for a set of gauge configurations 
%in the topologically-trivial sector 
with $ Q_{t} = 0 $, it gives $ \chi_t = 0 $.
However, even for a topologically-trivial gauge configuration,
it may possess non-trivial topological excitations in sub-volumes.
Thus, one can measure $ \chi_t $
using the correlation of the topological charges of two sub-volumes.  

In general, for any topological sector with $ Q_t $,
using saddle point expansion on the QCD partition function
in a finite volume, it can be shown that \cite{Aoki:2007ka}
\bea
\label{eq:rho_rho}
\lim_{|x| \to \infty} \left< \rho(x) \rho(0) \right> =
\frac{1}{\vol} \left( \frac{Q_t^2}{\vol} - \chi_t
                     -\frac{c_4}{2 \chi_t \vol} \right)
 + {\cal O}(\vol^{-3}), 
\eea
where
$
c_4 = -\frac{1}{\vol} \left[   \langle Q_t^4 \rangle_{\theta=0}
                            -3 \langle Q_t^2 \rangle_{\theta=0}^2 \right]
$.
However, for lattice QCD, it is difficult to extract $ \rho(x) $ and $ Q_t $
unambiguously from the gauge link variables, due to their rather
strong fluctuations.

To circumvent this difficulty, one may consider
the Atiyah-Singer index theorem
\cite{Atiyah:1968mp}
\bea
\label{eq:AS_thm}
Q_t = n_+ - n_- = \mbox{index}({\cal D}), 
\eea
where $ n_\pm $ is the number of zero modes of the massless Dirac
operator $ {\cal D} \equiv \gamma_\mu ( \partial_\mu + i g A_\mu) $
with $ \pm $ chirality. Since $ {\cal D} $ is anti-Hermitian and chirally
symmetric, its nonzero eigenmodes must come in complex conjugate pairs
(i.e., $ {\cal D} \phi = i \lambda \phi $ implies
       $ {\cal D} \gamma_5 \phi = -i \lambda \gamma_5 \phi $, 
for $ \lambda = \lambda^* \ne 0 $)
with zero chirality ($ \int d^4 x \phi^{\dagger} \gamma_5 \phi = 0 $).
Thus one can obtain the identity
\bea
\label{eq:index_Q1}
n_+ - n_- = \int d^4 x \ m~\tr [ \gamma_5 ({\cal D} + m)^{-1}(x,x)], 
\eea
by spectral decomposition, where the nonzero modes drop out
due to zero chirality. In view of (\ref{eq:AS_thm}) and (\ref{eq:index_Q1}),
one can regard $ m~\tr [ \gamma_5 ({\cal D} + m)^{-1}(x,x)] $
as topological charge density, to replace $ \rho(x) $
in the measurement of $ \chi_t $.

For lattice QCD, it is well-known that the overlap Dirac operator
\cite{Neuberger:1997fp,Narayanan:1995gw}
in a topologically non-trivial gauge background
possesses exact zero modes (with definite chirality) satisfying
the Atiyah-Singer index theorem. Writing the massive 
overlap Dirac operator as
\BAN
D(m) = \left( m_0 + \frac{m}{2} \right) +
\left( m_0 - \frac{m}{2} \right) \gamma_5 \frac{H_w(-m_0)}{\sqrt{H_w^2(-m_0)}}, 
\EAN
where $ H_w(-m_0) $ is the standard Hermitian Wilson operator with negative
mass $ -m_0 $ ($ 0 < m_0 < 2 $), then
the topological charge density can be defined as
\BAN
\rho_m(x) = m~\tr[ \gamma_5 ( D_c + m)^{-1}_{x,x}], 
\EAN
where $ (D_c + m )^{-1} $ is the valence quark propagator with quark
mass $ m $, and $ D_c $ is a chirally symmetric operator 
relating to $ D(0) $ by $ D_c = D(0) [1 - D(0)/(2m_0)]^{-1} $ 
\cite{Chiu:1998gp}.
%Note that $ (D_c + m)^{-1} $ is exponentially-local
%for sufficiently smooth gauge background and nonzero $ m $. 
%Also $ D_c $ is well-defined (without any poles)
%for topologically-trivial gauge configuration.
Here $ \rho_m(x) $ is justified to be topological charge density,
since it can be shown that 
$ \sum_x \rho_m(x) = n_+ - n_-$, 
which is similar to its counterpart in continuum, (\ref{eq:index_Q1}). 

Now we can replace $ \rho(x) $ with $ \rho_1(x) $, and use
(\ref{eq:rho_rho}) to extract $ \chi_t $ for any topological sector.
However, on a finite lattice,
it is contaminated by $ m_\pi $, $ m_{\eta'} $ and any states which
can couple to $ \langle \rho_1(x) \rho_1(0) \rangle $.
An alternative is to consider the correlator of the flavor-singlet
pseudoscalar meson $ \eta' $ \cite{Aoki:2007ka}
\bea
\label{eq:etap_2pt}
 \lim_{|x_1 - x_2| \gg 1 } m_q^2
                   \left< \eta'(x_1) \eta'(x_2) \right>_Q
&=& -\frac{\chi_t}{\vol} \left( 1 - \frac{Q^2}{\chi_t\vol}
             + \frac{c_4}{2 \chi_t^2 \vol} \right) 
             + {\cal O}( e^{-m_{\eta'} |x_1 - x_2|})
             + {\cal O}(\vol^{-3}), 
\eea
which is equal to the disconnected part 
$ \left< \rho_1(x_1) \rho_1(x_2) \right>_Q $ at large separation, 
but it tends to the asymptotic value faster than the later since 
it only couples to the states containing $ \eta' $.   
Then the time-correlation function of $ \eta' $
is fitted to $ A + B( e^{-M t} + e^{-M(T-t)} ) $
to obtain the constant
$ A = \frac{1}{m_q^2} \frac{1}{T}
    \left( \frac{Q_t^2}{\vol} - \chi_t -
           \frac{c_4}{2 \chi_t \vol} \right) $, 
and from which to extract $ \chi_t $ 
provided that $ |c_4| \ll 2 \chi_t^2 \vol $.  
This was how we determined the topological susceptibility 
in two-flavor lattice QCD with fixed topology      
\cite{Chiu:2007hb,Aoki:2007pw}.

However, it was unclear to what extent the assumption  
$ |c_4| \ll 2 \chi_t^2 \vol $ was satisfied.
To eliminate this constraint, we compute the 4-point correlator 
of $ \eta' $, as well as the 2-point correlator. Theoretically,  
in a fixed topology, the former behaves as \cite{Aoki:2007ka}
\bea
\label{eq:etap_4pt}
\lim_{|x_i - x_j| \gg 1 } m_q^4
%\langle\eta'(x_1)\eta'(x_2)\eta'(x_3)\eta'(x_4)\rangle_Q
\langle\eta'(x_1) \cdots \eta'(x_4)\rangle_Q
= \frac{3\chi_t^2}{\vol^2}
  \left(1-\frac{Q^2}{\chi_t\vol} +\frac{c_4}{\chi_t^2 \vol} \right)^2
         + {\cal O}( e^{- m_{\eta'} |x_i-x_j|})
         + {\cal O}(\vol^{-4}),  
\eea
From (\ref{eq:etap_2pt}) and (\ref{eq:etap_4pt}), one can solve
for $ \chi_t $ and $ c_4 $ (or equivalently, the parameter $ y $)
\bea
\label{eq:chit}
\chi_t &=& \frac{Q^2}{\vol} + \vol \left( 2 k_2 - \sqrt{k_4/3} \right), 
\\
\label{eq:y}
y &\equiv& \frac{c_4}{2\chi_t^2 \vol}
= - \frac{\left(\sqrt{k_4/3} - k_2 \right)}{\sqrt{k_4/3} - 2 k_2}
        \left(1-\frac{Q^2}{\chi_t \vol} \right), 
\eea
where $ -k_2 $ and $ k_4 $ are the asymptotic values of 2-point and 4-point 
correlators at large separation. 
It is interesting to note that if one neglects the $ y $ term
in (\ref{eq:etap_2pt}) and (\ref{eq:etap_4pt}), they reduce to
\bea
\label{eq:chit_2pt}
\chi_t &\simeq& \frac{Q^2}{\vol} + \vol k_2,     \\
\label{eq:chit_4pt}
\chi_t &\simeq& \frac{Q^2}{\vol} + \vol \sqrt{k_4/3}, 
\eea
which provide two independent estimates of $ \chi_t $.
In other words, if $ |y| \ll 1 $, then (\ref{eq:chit}), (\ref{eq:chit_2pt})
and (\ref{eq:chit_4pt}) all give compatible results for $ \chi_t $.
On the other hand, if (\ref{eq:chit_2pt}) and (\ref{eq:chit_4pt})
turn out to be quite different from each other,
then $ |y| $ must be substantially larger than zero, 
and a more reliable estimate of $ \chi_t $ 
could be given by (\ref{eq:chit_4pt}). 

For the (2+1)-flavor QCD, the $ \eta' $ interpolating operator must 
take into account of the fact that different flavors have 
different quark masses, namely, 
\bea
m_q \eta' = \frac{m_q}{N_f} \sum_{f=1}^{N_f} \bar q_f \gamma_5 q_f 
& \longrightarrow &   
\eta'_T = \frac{1}{N_f} \sum_{f=1}^{N_f} m_f \bar q_f \gamma_5 q_f 
\eea
where $ \eta'_T $ is called the ``topological" $\eta'$ operator for 
computing topological charge density correlators. 

In this paper, we use 80 pairs of low-lying eigenmodes of the overlap 
operator to evaluate the 2-point and 4-point correlators of $ \eta'_T $, and
to extract their asympototic values $ -k_2 $ and $ k_4 $. 
Then we use (\ref{eq:chit})-(\ref{eq:y}) to obtain $ \chi_t $ and $ y $.
Note that $ c_4 $ is related to the leading anomalous contribution to 
the $ \eta'-\eta' $ scattering amplitude in QCD, as well as the 
dependence of the vacuum energy on the vacuum angle $ \theta $.     

\begin{figure}[htb]
\begin{center}
\begin{tabular}{@{}cc@{}}
\includegraphics*[height=5cm,width=6cm]{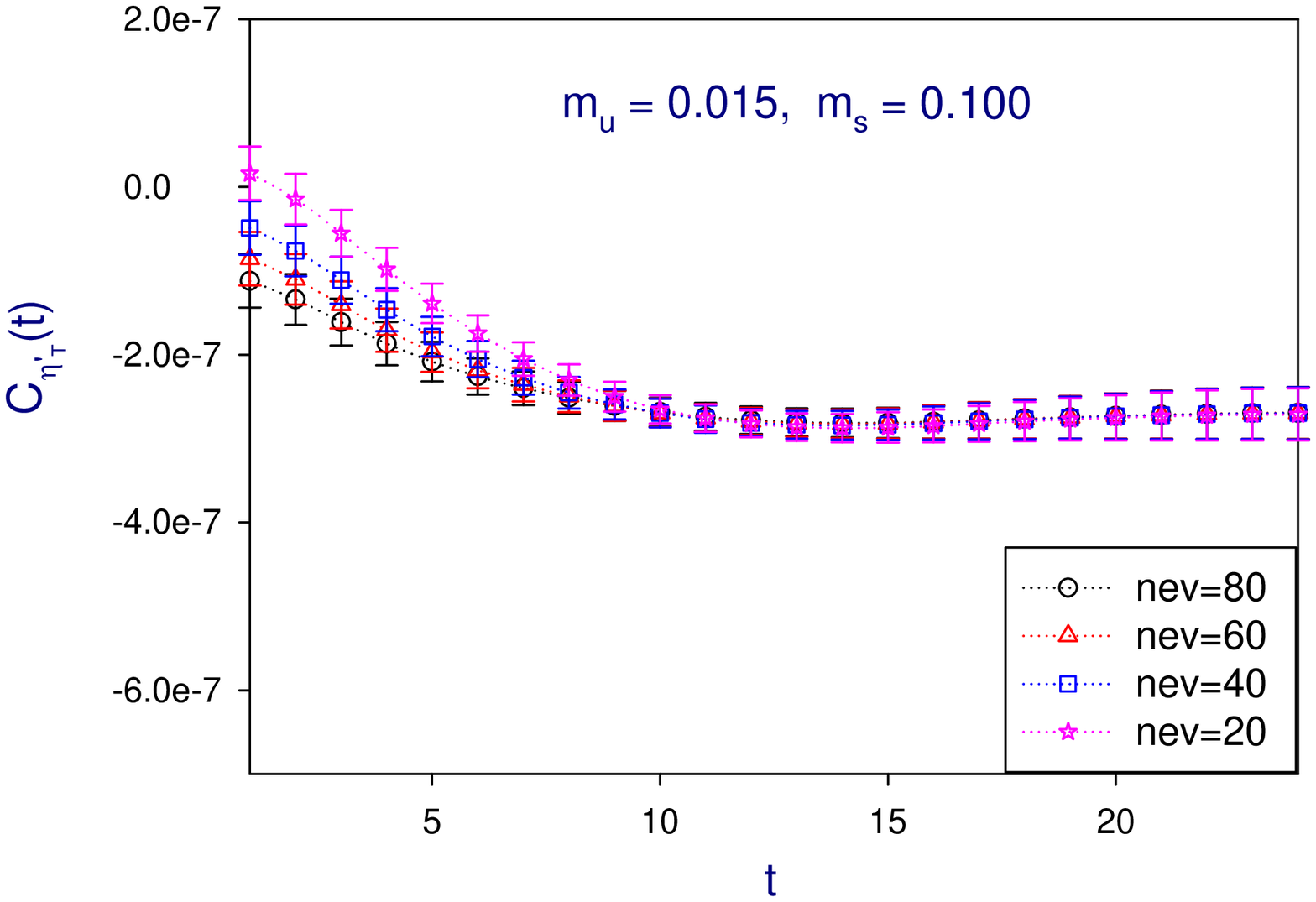}
&
\includegraphics*[height=5cm,width=6cm]{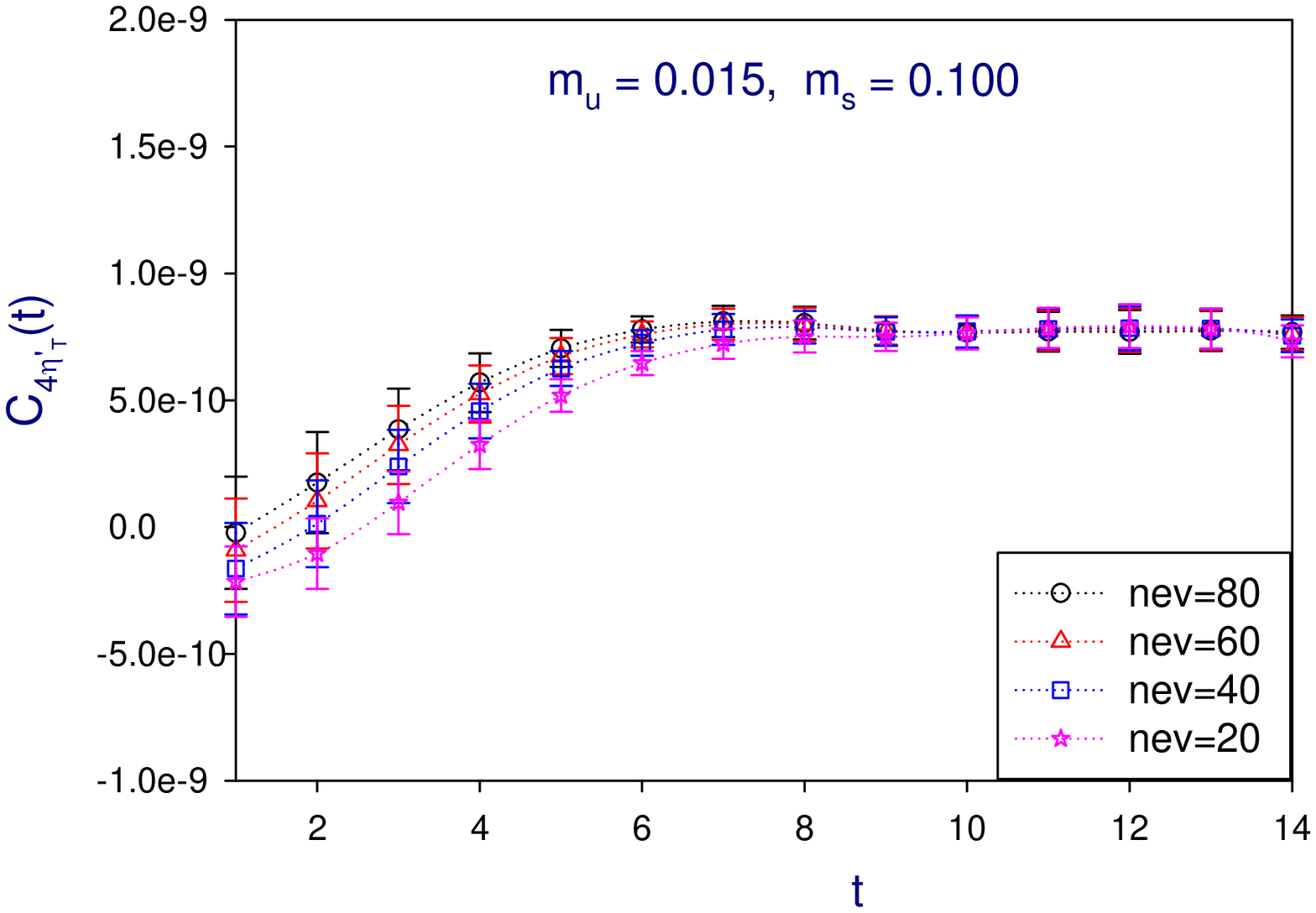}
\\ (a) & (b)
\end{tabular}
\caption{ Low-mode saturation of 
          (a) the 2-point function $ C_{\eta'_T}(t) $
          (b) the 4-point function $ C_{4\eta'_T}(t) $
}
\label{fig:etap_2pt_4pt}
\end{center}
\end{figure}

\section{Lattice Setup}

Our simulations are carried out in the topologically-trivial sector 
(with $ Q_t = 0 $) for (2+1)-flavor QCD 
on a $16^3\times 48$ lattice at a lattice spacing $\sim$ 0.11~fm
(for details, see \cite{Hashimoto:2007vv} and \cite{Matsufuru:2008}).
For the gluon part, the Iwasaki action is used at $\beta$ = 2.30, 
together with unphysical Wilson fermions and associated twisted-mass ghosts
\cite{Fukaya:2006vs}. The unphysical degrees of freedom generate a factor 
$\det[H_w^2(-m_0)/(H_w^2(-m_0)+\mu^2)]$ in the partition function 
(we take $m_0=1.6$ and $\mu=0.2$)
that suppresses the near-zero eigenvalue of $H_w(-m_0)$ and thus
makes the numerical operation with the overlap operator  
substantially faster. 
Furthermore, since the exact zero eigenvalue is forbidden, the global
topological change is preserved  
during the molecular dynamics evolution of the gauge field.  

For $ m_s = 0.100$, we take five sea quark mass $m_{u(d)}$ 
values: 0.015, 0.025, 0.035, 0.050, and 0.100 
that cover the mass range $m_s/6$--$m_s$.
After discarding 500 trajectories for thermalization, we accumulate 
2500 trajectories in total for each sea quark mass.
In the calculation of $\chi_t$, we take one configuration every 5
trajectories, thus we have 500 configurations for each $ m_q $.
For each configuration, 80 pairs of lowest-lying eigenmodes 
of the overlap-Dirac operator $D(0)$ are calculated using the
implicitly restarted Lanczos algorithm and stored for the later use.

\section{Results}

In practice, we use 80 pairs of low-lying eigenmodes of the overlap operator
to evaluate the 2-point and 4-point time-correlation functions
of $\eta'_T$
\BAN
C_{\eta'_T}(t) &=& \frac{1}{L^3T}
                  \sum_{u=1}^T \sum_{\vec x_i}
                  \left< \eta'_T(\vec x_2,u+t)
                         \eta'_T(\vec x_1,u) \right>, 
\hspace{4mm} 
\lim_{t \gg 1} \frac{1}{L^3} C_{\eta'_T}(t) = -k_2, 
\\
C_{4\eta'_T}(t) &=& \frac{1}{L^3T}
                  \sum_{u=1}^T \sum_{\vec x_i}
                  \left< \eta'_T(\vec x_4,u+3t)
                         \eta'_T(\vec x_3,u+2t)
                         \eta'_T(\vec x_2,u+t)
                         \eta'_T(\vec x_1,u)  \right>, 
\hspace{4mm} 
\lim_{t \gg 1} \frac{1}{L^9} C_{4\eta'_T}(t) = k_4.
\EAN
Thus it is crucial to check whether these 80 eigenmodes suffice to 
saturate $C_{\eta'_T}(t)$ and $C_{4\eta'_T}(t)$ respectively. 
In Fig. \ref{fig:etap_2pt_4pt}, we plot  
$C_{\eta'_T}(t)$ and $C_{4\eta'_T}(t)$ for $ m_u = 0.015 $, 
versus the number of eigenmodes (nev) 20, 40, 60, and 80 respectively. 
Obviously, $ C_{\eta'_T}(t)$ is well saturated with 80 eigenmodes 
for the time range $15 \le t \le 24 $ where it attains a plateau.   
Similarly, $ C_{4\eta'_T}(t)$ is also well saturated  
for the time range $9 \le t \le 14$ where it attains a plateau.   
The low-mode saturation also holds for all five sea quark masses. 
 
\begin{figure}[tb]
\begin{center}
\begin{tabular}{@{}cc@{}}
\includegraphics*[height=5cm,width=6cm]{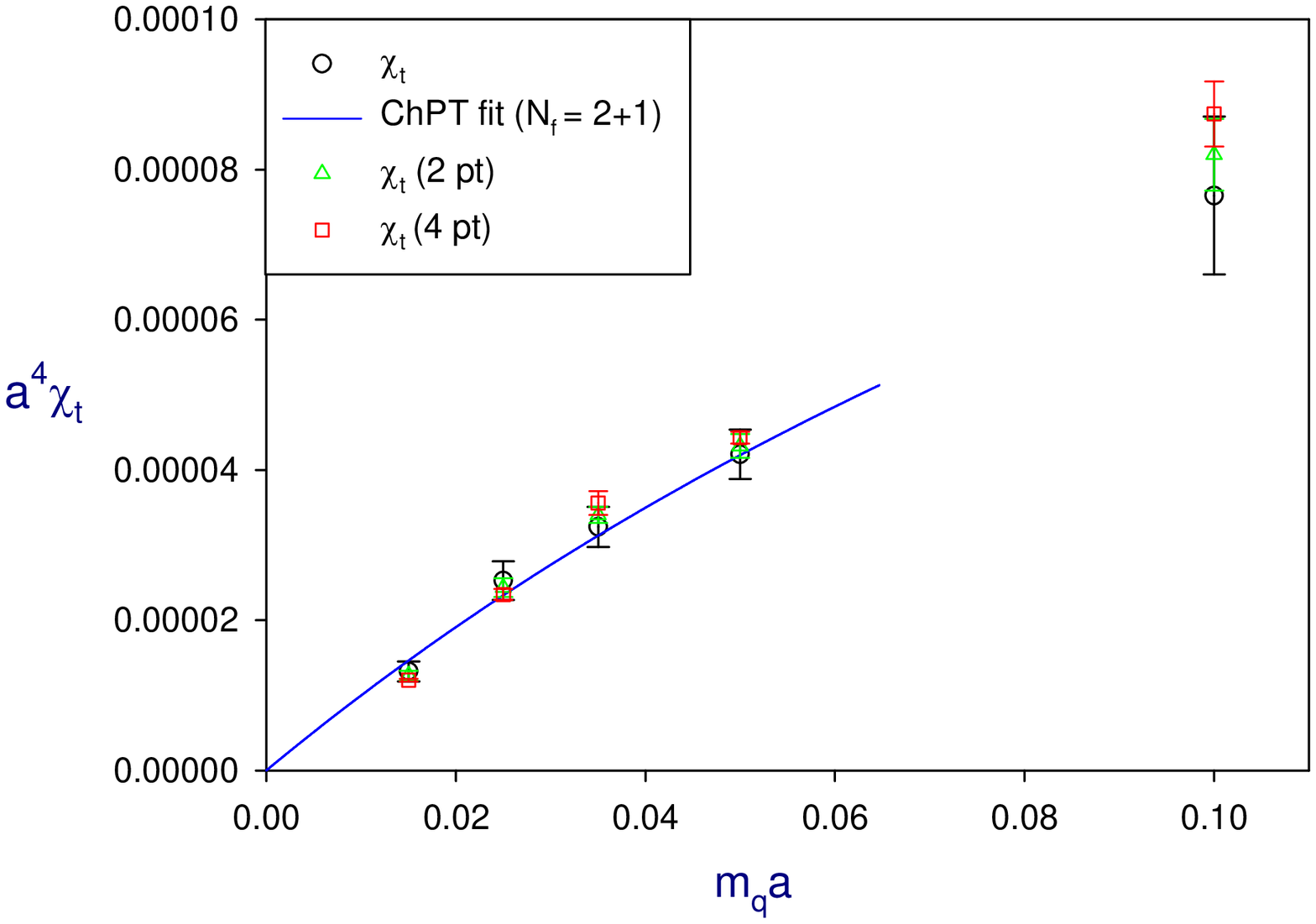}
&
\includegraphics*[height=5cm,width=6cm]{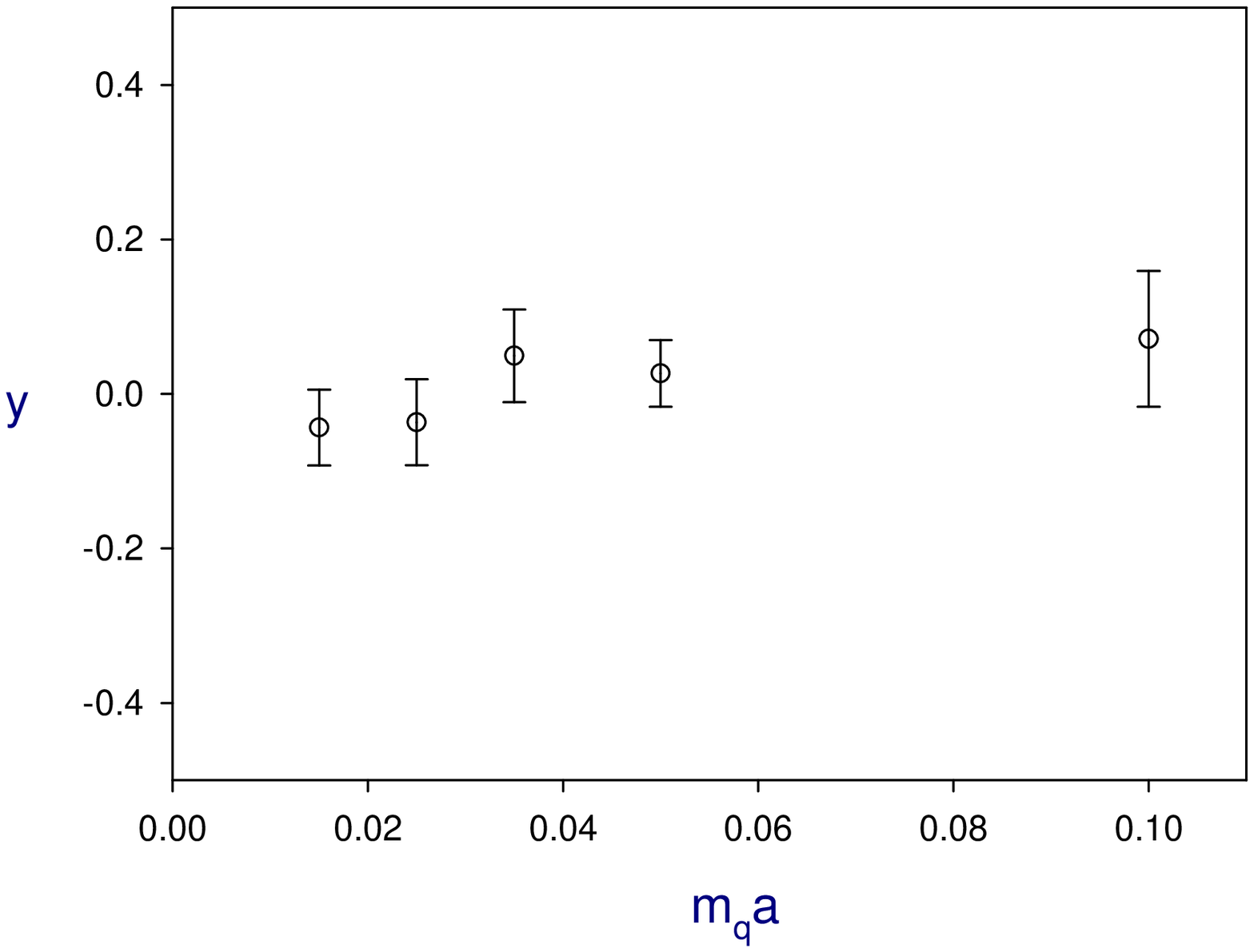}
\\ (a) & (b)
\end{tabular}
\caption{
    Topological susceptibility $ \chi_t a^4 $ and 
    $ y \equiv c_4/(2\chi_t^2 \vol) $
    versus sea quark mass $m_q a$ for (2+1)-flavor lattice QCD with 
    fixed topological charge $ Q_t = 0 $.}
\label{fig:chit_y_mq_Q0_ms01}
\end{center}
\end{figure}

In Fig.~\ref{fig:chit_y_mq_Q0_ms01}, we plot the values of  
$ \chi_t a^4 $ (\ref{eq:chit}) and $ y $ (\ref{eq:y}) 
versus the sea quark mass $ m_q a $, together with the values 
of $ \chi_t $ obtained from the 2-point function (\ref{eq:chit_2pt})   
and the 4-point function (\ref{eq:chit_4pt}) respectively. 
Evidently, the values of $ \chi_t $ from (\ref{eq:chit}),  
(\ref{eq:chit_2pt}), and (\ref{eq:chit_4pt}) are in good agreement
with one another. 
%At $ m_q a = 0.10 $, both $ k_2 $ and $ k_4 $ have large errors,
%which in turn give $ y = 0.071(88) $, consistent with zero but 
%with a much larger error than those of other sea quark masses.  
%  
For the smallest four quark masses, 0.015, 0.025, 0.035, and 0.050, 
the data points of $ a^4 \chi_t $ are well fitted by  
the ChPT formula \cite{Leutwyler:1992yt} 
\bea
\label{eq:ChPT_chit}
\chi_t = \frac{\Sigma}{m_u^{-1} + m_d^{-1} + m_s^{-1}}, 
\eea
with $ a^3 \Sigma = 0.0021(1) $. 
In order to convert $\Sigma$ to that in the
$\overline{\mathrm{MS}}$ scheme, we calculate the 
renormalization factor $Z_m^{\overline{\mathrm{MS}}}(\mathrm{2~GeV})$
using the non-perturbative renormalization technique
through the RI/MOM scheme. Our result is 
$Z_m^{\overline{{\mathrm{MS}}}}(\{mathrm{2~GeV}) = 0.833(8)$ 
\cite{Noaki:2008gx}. 
With $ a^{-1} = 1833(12)$ MeV determined with $ r_0 = 0.49 $ fm 
\cite{Matsufuru:2008}, the value of $ \Sigma $ is transcribed to
$$
\Sigma^{\overline{{\mathrm{MS}}}}(\mathrm{2~GeV})
  =[249(4)(2) \mathrm{MeV}]^3, 
$$ 
which is in good agreement that extracted from 
$ \chi_t = \langle Q_t^2 \rangle / \vol $ 
with $ Q_t $ determined by the spectral flow method   
for the 2+1 flavors QCD configurations generated by the 
RBC and UKQCD Collaborations with domain-wall fermions \cite{Chiu:2008jq}.
Also, it is in good agreement with our previous results 
extracted from $ \chi_t $ in 2-flavor QCD \cite{Chiu:2007hb,Aoki:2007pw}, 
and in the $\epsilon$-regime from 
the low-lying eigenvalues \cite{Fukaya:2007yv}.
%\cite{Fukaya:2007fb,Fukaya:2007yv}. 
The errors represent a combined statistical error
($a^{-1}$ and $Z_m^{\overline{\mathrm{MS}}}$) and
the systematic error respectively.
%Since the calculation is done at a single lattice spacing,
%the discretization error cannot be quantified reliably, but
%we do not expect much larger error because our lattice
%action is free from $O(a)$ discretization effects.

\begin{figure}[tb]
\centering
\includegraphics[width=8cm,height=6cm,clip=true]
                {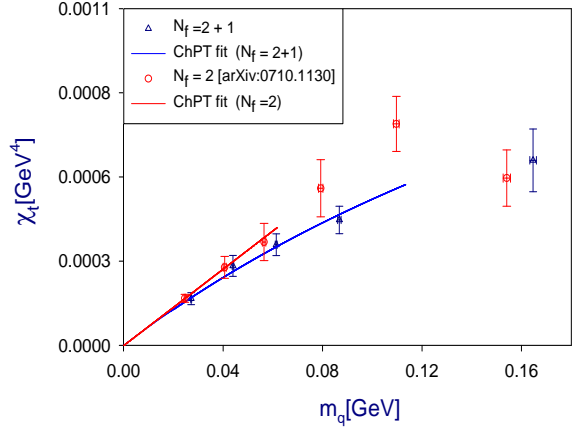}
\label{fig:chit_mq_Q0_nf2_nf2p1ms01}
\caption{The topological susceptibility $ \chi_t $
         versus $ m_q $ for lattice QCD with fixed topology $ Q_t = 0 $.}
\end{figure}

At this point, it is instructive to plot $ \chi_t $ versus
$ m_q $, for 2-flavor QCD (data from \cite{Chiu:2007hb,Aoki:2007pw}), 
and (2+1)-flavor QCD (this work),  
as shown in Fig. \ref{fig:chit_mq_Q0_nf2_nf2p1ms01}. 
Now we can see clearly how the topological susceptibility 
changes with respect to the number of flavors.   

\section{Concluding remark}
  
In this paper, we have obtained the topological susceptibility $\chi_t$
and $ c_4 $ in (2+1)-flavor QCD from 
a lattice calculation of 2-point and 4-point correlators at 
a fixed global topological charge $ Q_t = 0 $. 
The expected sea quark mass dependence of $\chi_t$ from ChPT is clearly
observed. 
Our result asserts that the topologically non-trivial excitations 
are in fact locally active in the QCD vacuum, even when the global 
topological charge is zero. 
The information of these topological excitations is carried by the low-lying
eigenmodes of the overlap Dirac operator. 
We will use the values of $ \chi_t $ we have determined  
to remove the artifacts due to the fixed topology in a finite volume 
and to obtain the physical results in the $\theta$ vacuum 
\cite{Brower:2003yx, Aoki:2007ka}. 

Finally we note that our result of the ratio $ |c_4|/\chi_t $ is 
substantially less than one in the chiral limit, similar to 
its counterpart in quenched QCD \cite{Giusti:2007tu,DelDebbio:2007kz}.
This seems to suggest that the quantum corrections would suppress the 
emergence of dilute instanton gas in the full QCD vacuum.  
% $ F(\theta) \propto -\vol[\cos(\theta)-1] $
% predicted by the dilute instanton model 
% is disfavored. 

%\begin{acknowledgments}
  Numerical simulations are performed on Hitachi SR11000 and IBM System Blue
  Gene Solution at High Energy Accelerator Research Organization (KEK) under 
  a support of its Large Scale Simulation Program (No.~08-05), 
  and also on IBM and HP clusters at NCHC and NTU-CC in Taiwan.
  This work is supported in part by the Grant-in-Aid of the
  Japanese Ministry of Education 
  (Nos.~18340075,  % Hashimoto   
       18740167,   % Yamada     
       19540286,   % Noaki
       19740160,   % Matsufuru
       20025010,   % Yamada
       20039005,   % Onogi
       20340047,   % Aoki
       20740156),  % Onogi 
   the National Science Council of Taiwan 
   (Nos.~NSC96-2112-M-002-020-MY3, 
        NSC96-2112-M-001-017-MY3, 
        NSC97-2119-M-002-001),   
    and NTU-CQSE (Nos.~97R0066-65, 97R0066-69). 
%\end{acknowledgments}

\end{document}